\documentstyle[amssymb,aps,prd,epsf]{revtex}

\begin{document}
\author{E. Elizalde$^{1,2,}$\footnote{Presently on leave at:
Department of Mathematics,  Massachusetts Institute
of Technology, 77 Massachusetts Avenue, Cambridge, MA 02139-4307;
elizalde@math.mit.edu; elizalde@ieec.fcr.es},
E. J. Ferrer$^{3}$ \and and V. de la Incera$^{3}$}
\address{$^{1}$Institute for Space Studies of Catalonia, IEEC/CSIC,\\
 Edifici Nexus, Gran Capit\`a 2-4, 08034 Barcelona, Spain.\\
$^{2}$Dept. of Structure and Constituents
of Matter, Univ. of Barcelona,  \\
Diagonal 647, 08028 Barcelona, Spain. \\
$^{3}$ Physics Dept., State University of New York at Fredonia,\\
 Houghton Hall, Fredonia, NY 14063, USA}

\title{Beyond-Constant-Mass-Approximation Magnetic Catalysis in
the Gauge Higgs-Yukawa Model} \maketitle
\date{}

\begin{abstract}
Beyond-constant-mass approximation solutions for magnetically
catalyzed fermion and scalar masses are found in a gauge
Higgs-Yukawa theory in the presence of a constant magnetic field.
The obtained fermion masses are several orders of magnitude larger
than those found in the absence of Yukawa interactions. The masses
obtained within the beyond-constant-mass approximation exactly
reduce to the results within the constant-mass approach when the
condition $\nu \ln (\frac{1}{\widehat{m}^{2}})\ll 1$ is satisfied.
Possible applications to early universe physics and condensed
matter are discussed. \vskip1mm PACS numbers: 11.30Qc; 11.30 Rd;
12.15-y; 12.20.Ds
\end{abstract}

\section{Introduction}

In the last few years the magnetic catalysis (MC) of chiral
symmetry breaking \cite{mira-gus-sho}-\cite{Klimenko} has been the
focus of attention of many works on non-perturbative effects of
magnetic fields \cite{mira-gus-sho}-\cite{ferrer-incera5}. The
phenomenon consists on the dynamical generation of a fermion
condensate (and consequently of a fermion mass) when the fermion
interactions occur in the presence of an external constant
magnetic field. A most significant feature of the MC is that it
requires no critical value of the fermion's coupling for the
condensate to be generated. That is, the symmetry breaking takes
place at the weakest attractive interaction. Physically it is due
to the fact that the magnetic field forces the low energy fermions
to reside basically in their lowest Landau level (LLL), while the
higher energy fermions actually decouple \cite{hong2}. This, in
turn, yields a dimensional reduction of the infrared fermion
dynamics. The dimensional reduction is reflected in an effective
strengthening of the fermion interactions leading to dynamical
symmetry breaking through the generation of a fermion condensate.

A particularly important question to understand in this context is
how the MC is affected by the introduction of fermion-scalar
interactions. Fermion-scalar interactions are an essential element
of the unified theories of fundamental forces. As is well known,
they are expected to be responsible for the fermion mass appearing
due to the spontaneous symmetry breaking of the electroweak
symmetry. Fermion-scalar interactions are also relevant in
condensed matter physics where the complexity of
strongly-correlated many-body systems some times calls for a
description in terms of more simple, phenomenological theories
that contain interacting scalar in addition to fermions (see e.g.
\cite{vojta}).

In Refs. \cite{ferrer-incera2}-\cite{ferrer-incera1} two of us
studied the realization of magnetic catalysis in a
(3+1)-dimensional Higgs-Yukawa (HY) model, showing that the
magnetic-field-induced fermion mass is enhanced by fermion-scalar
interactions. As we will show below, this enhancement is also
found within a more accurate approximation for a wide range of
couplings. This result might find applications in early universe
transitions, as well as in condensed matter physics.

In \cite{ferrer-incera2}-\cite{incera} some applications of the MC
to the early universe were briefly considered. They were motivated
by many astrophysical observations of galactic and intergalactic
magnetic fields indicating the existence of seed fields that
originated from large primordial magnetic fields (for a recent
review on cosmic magnetic fields see \cite{grasso}). If the
primordial magnetic fields in the early universe were large
compared to the values close to the phase transition point of the
fermion masses generated through the usual mechanism of
spontaneous symmetry breaking, the fermion would seem
approximately massless. Under these circumstances, it is important
to investigate if the primordial magnetic fields could contribute
to the masses of the fermions through MC and hence influence the
phenomenology of the early universe \cite{ferrer-incera1}.

On the other hand, to discuss applications of MC in the context of
a HY theory to condensed matter, we need, besides interactions
modelled by fermion-scalar terms, a physical system that, despite
being non-relativistic, can be described under certain conditions
by a "relativistic" Hamiltonian. We will see below that these
conditions are indeed present in the physics of high-$T_{c}$
superconductors.

High-$T_{c}$ superconductors, which are characterized by the
existence of nodal points where the order parameter (gap function)
vanishes, provide a practical realization of a "relativistic"
system in condensed matter physics. This is so because the
low-energy spectrum of the nodal quasiparticles is linear, hence
the quasiparticle excitations are described by an anisotropic
Dirac Hamiltonian \cite{durst-lee}. In Ref. \cite{vojta} a
quantum-critical phase transition to a new superconducting state,
characterized by the appearance of a secondary pairing at some
doping level, was proposed to explain recent measurements
\cite{exp-1} of an anomalously large inelastic scattering of
quasiparticles near the gap nodes of a superconductor. The
observed secondary pairing transition made the nodal
quasiparticles fully gapped. Based on the symmetries of the
superconductor, the authors of Ref. \cite{vojta} made a
classification of a set of fermion-scalar interactions that in
principle could be in agreement with the experimental
observations, and then performed a perturbative
renormalization-group analysis of each model to determine the
possible existence of a quantum-critical point. In Ref.
\cite{kvesh}, expanding on the ideas of \cite{vojta}, the
existence of a quantum critical point was established directly in
a (2+1)-dimensional HY theory beyond a non-perturbative approach,
which allowed to make quantitative predictions for the
corresponding quantum-critical behavior. The gap generation
(fermion mass) was associated in \cite{kvesh} with the breaking of
a discrete chiral symmetry.

We would like to underline that the breaking of the chiral
symmetry in \cite{kvesh} was found to occur when the Yukawa
coupling (assumed to be related to the doping level) reached a
critical value, that is, the symmetry breaking was not associated
to the phenomenon of MC, as no external magnetic field was
introduced in the analysis. However, as recently observed
\cite{dagan} by measuring the splitting of the conductance peak
that characterizes the nodes of high-$T_{c}$ superconductors, the
development of a secondary quasiparticle gap may be triggered not
only by the doping level, but also by an applied magnetic field.
Could the secondary gap triggered by the magnetic field be the
consequence of MC occurring within the superconductor? We believe
that the results we are going to derive below strongly indicate
that the answer is yes, if, as argued in \cite{vojta} and
\cite{kvesh}, the HY theory is the model describing the appearing
of the secondary gap. Nevertheless, to match the experimental
observations we would need to particularize the analysis done in
the present paper to the (2+1)-dimensional case, and adjust the
physical values of the couplings to those characteristic of the
superconductor.

As already mentioned, in Ref. \cite{ferrer-incera1} the phenomenon
of MC in a (3+1)-dimensional Abelian gauge theory with HY
interactions was studied. In that work it was shown that the
non-perturbative solution of the minimum equations for the
composite-operator effective action leads not only to a
magnetically catalyzed fermion dynamical mass, but also to a
nonzero scalar vev $\varphi_{c}$ and consequently, to a nonzero
scalar mass. In other words, thanks to the magnetic field, a
scalar-field minimum solution is generated by non-perturbative
radiative corrections.

We should underline though that the fermion and scalar masses of
ref. \cite{ferrer-incera1} were obtained within a simplified
approximation known in the literature as the constant mass
approximation (CMA). In general, to find the dynamical mass
---which is nothing but the part of the fermion self-energy
proportional to the identity matrix--- one has to solve a
non-perturbative gap equation (i.e. the Schwinger-Dyson equation
for the full fermion propagator). This means to solve a
non-linear, implicit integral equation for the fermion
self-energy, which is a momentum-dependent function. Most authors
approach such a mathematically complicated problem with the help
of the rough CMA approach. It consists on neglecting the momentum
dependence of the self-energy in the gap equation. This is done by
substituting the self-energy function in the gap equation by its
value at zero momentum, that is, by the infrared mass. There is no
general principle that guarantees the validity of this
approximation for the whole range of physical couplings.

For theories with several couplings, due to the richness of the
parameter space, the reliability of the CMA is questionable and
should be investigated in detail.  In the case of the HY model,
aside from the multiple-coupling problem, one has to deal with a
system of non-linear, coupled integral equations, one for the
fermion dynamical mass and other for the scalar vev
\cite{ferrer-incera1}. We cannot disregard in this situation the
possibility of regions of these parameters where the CMA is
reliable and regions where it is not. In this case one has to turn
to a more accurate approximation on which the momentum dependence
of the self-energy is taken into account when solving the gap
equation. This more accurate approximation is known as the
beyond-constant-mass approximation (BCMA).

For theories like QED containing only one coupling constant, the
CMA is known to be appropriate, since going beyond it does not
produce qualitatively different results. This has been explicitly
shown for (3+1)-dimensional \cite{Gusynin} and (2+1)-dimensional
QED \cite{alexandre}, and independently corroborated by our
calculations below. In Ref. \cite{gms2} the BCMA mass solution of
(3+1)-dimensional QED was found to agree with the CMA mass
obtained from the improved-ladder\footnote{One should distinguish
between the approximation employed to obtain the gap equation
itself (ladder, improved ladder, etc), and the approximation
employed to find its solution. In the present paper we are
concerned with the approximation to find the mass solution (either
CMA or BCMA), assuming that the gap equation is found using a
ladder or an improved ladder approximation.} gap equation. This
result was later corroborated by numerical calculations in
\cite{alexandre2}.

Considering that different physical applications of the MC in
models with fermion-scalar interactions would require different
values of the couplings constants, and in particular, given the
relevance that the Abelian gauge Higgs-Yukawa theory may have for
condensed matter and other field theory applications, it is
important to perform a BCMA investigation of this model in all
possible regions of the parameter space and find out whether it
significantly differs or not from the CMA results. A main goal of
the present paper is to carry out such a study.

By going beyond the CMA, we will determine the region of Yukawa
and scalar self-interaction couplings where the CMA is valid, and
will obtain the numerical BCMA solutions for the fermions and
scalar dynamical masses in the complete physically meaningful
parameter region. As we will see below, the CMA results for the
Abelian gauge Higgs-Yukawa theory are mostly reliable in the
available parameter space. An important finding is that the
(BCMA-found) mass values are many orders of magnitude larger than
those obtained in the absence of fermion-scalar interactions,
corroborating, within this more accurate approximation, the
enhancement of the dynamical mass by the Yukawa term.

The paper is organized as follows: In Section II we derive the
non-linear integral equations for the fermion self energy (gap
equation) and the scalar vev in a gauge Higgs-Yukawa theory. The
integral gap equation is then converted into a second order
differential equation with boundary conditions. In Section III,
this boundary-value problem is analytically solved leading to the
self energy as a function of the momentum and the infrared fermion
mass. Using the self-energy solution and the equation for the
scalar minimum, we arrive at two coupled transcendental equations
depending on the infrared dynamical mass and the scalar vev. These
equations are numerically solved and the results are used to
determine the region of reliability of the CMA and to compare the
mass values obtained in the CMA and in the BCMA approaches. We end
Section III discussing the solution of the gap equation at zero
Yukawa coupling, and showing that it leads to the same result
found in Ref.\cite{Gusynin} for (3+1)-dimensional QED. In Section
IV, we state our concluding remarks and reconsider the question of
the relevance of the magnetic catalysis in the electroweak phase
transition using the BCMA results.
\section{Integral Equations}

Let us consider the following Lagrangian density

\begin{equation}
L=-\frac{1}{4}F^{\mu \nu }F_{\mu \nu }+i\overline{\psi }\gamma
^{\mu
}\partial _{\mu }\psi +e\overline{\psi }\gamma ^{\mu }\psi A_{\mu }-\frac{1}{%
2}\partial _{\mu }\varphi \partial ^{\mu }\varphi -\frac{\lambda }{4!}%
\varphi ^{4}-\frac{\mu ^{2}}{2}\varphi ^{2}-\lambda _{y}\varphi \overline{%
\psi }\psi  \label{lag}
\end{equation}
that describes a gauge Higgs-Yukawa model with a fermion field
coupled to scalar and electromagnetic fields. The scalar field is
electrically neutral, but self-interacting.

The Lagrangian density (\ref{lag}) has U(1) gauge symmetry,
\begin{eqnarray}
A_{\mu } &\rightarrow &A_{\mu }+\frac{1}{e}\partial _{\mu }\alpha
(x)
\nonumber \\
\psi &\rightarrow &e^{i\alpha (x)}\psi ,  \label{sym1}
\end{eqnarray}
fermion number global symmetry
\begin{equation}
\psi \rightarrow e^{i\theta }\psi ,  \label{sym2}
\end{equation}
and discrete chiral symmetry
\begin{equation}
\psi \rightarrow \gamma _{_{5}}\psi ,\qquad \overline{\psi }\rightarrow -%
\overline{\psi }\gamma _{_{5}},\qquad \varphi \rightarrow -\varphi .
\label{sym3}
\end{equation}

Notice that the quadratic scalar term has the correct sign of a
mass term, thus no vacuum expectation value of the scalar field
exists at tree level. In the course of our calculations we will
take $\mu\rightarrow0$ to search for a dynamically induced mass.
The discrete symmetry (\ref{sym3}) forbids a mass for the fermions
to all orders in perturbation theory. Nevertheless, this symmetry
could be dynamically broken through non-perturbative generation of
a composite field (fermion-antifermion condensate). Such a fermion
condensate would lead to a dynamical fermion mass and to a
non-zero vacuum expectation value of the scalar field
\cite{ferrer-incera1}, which in turn would contribute to the
scalar mass.

It is known that in the case of the non-gauge (3+1)-dimensional
Higgs-Yukawa theory, no value exists for a running $\lambda_{y}$
at which a chiral symmetry breaking fermion condensate can be
generated\footnote[1]{The incorporation of gauge field terms in
the Higgs-Yukawa model may lead to chiral symmetry breaking at
some critical $\alpha$, just as it occurs in
(3+1)-QED\cite{mira-gus-sho}-\cite{Gusynin}.}. As shown in
\cite{ferrer-incera1}, the situation drastically changes when a
magnetic field is introduced. In this case a non-trivial solution
exists at the weakest value of $\lambda_{y}$ and one can show that
a fermion condensate (together with a dynamical fermion mass and a
scalar vev) is magnetically catalyzed.

However, as already mentioned, the solutions in
Ref.\cite{ferrer-incera1} were found within the CMA, and therefore
it is important to investigate their reliability beyond that
approximation. Our task hereafter will be to extend the results of
Ref.\cite{ferrer-incera1} beyond the CMA to find the dynamical
mass and the scalar vev for all physically meaningful values of
$\lambda_{y}$ and $\lambda$. For the sake of understanding, we
will repeat the outline of the derivations done in
Ref.\cite{ferrer-incera1} that lead to the coupled set of integral
equations (gap and scalar vev equations) that will be the starting
point of our new calculations.

Let us consider the Lagrangian density (\ref{lag}) in the presence
of an external constant magnetic field $B$ (without loss of
generality we assume that the magnetic field is directed along the
third coordinate axis and that
sgn $(eB)>0$), which can be introduced by adding the external potential $%
A_{\mu }=\left( 0,0,eBx_{1},0\right) $ as a shift to the oscillatory gauge
field $%
A_{\mu }$ in Eq. (\ref{lag}). To find the vacuum solutions of this
theory we need to solve the extremum equations of the effective
action $\Gamma $ for composite
operators\cite{jackiw},\cite{miransbook}

\begin{eqnarray}
\frac{\delta \Gamma _{B}(\varphi _{c},\overline{G})}{\delta
\overline{G}}
&=&0,\;  \label{min1} \\
\frac{\delta \Gamma _{B}(\varphi _{c},\overline{G})}{\delta
\varphi _{c}} &=&0  \label{min2}
\end{eqnarray}
In the above $\overline{G}(x,x)=\sigma (x)=\left\langle 0\mid
\overline{\psi }(x)\psi (x)\mid 0\right\rangle $ is a composite
fermion-antifermion field, and $\varphi _{c}$ represents the vev
of the scalar field. The subindex $\textit{B}$ indicates that the
effective action is considered in the background of the external
magnetic field.

Equations $\left( \ref{min1}\right) $ and $\left(
\ref{min2}\right) $ are, respectively, the Schwinger-Dyson (SD)
equation for the fermion self-energy operator $\Sigma $ (gap
equation) and the minimum equation for the vev of the scalar
field. As we are interested in the possibility of a scalar mass
induced ---through the interactions with the fermions--- by a
dynamically generated fermion condensate, we will set, as stated
above, the bare scalar mass $\mu $ to zero. Notice that, if the
minimum solutions of Eqs. (\ref{min1}) and (\ref{min2}) are non
trivial, the discrete chiral symmetry (\ref{sym3}) is dynamically
broken and both fermions and scalars acquire mass. The loop
expansion of the effective action $\Gamma $ for composite
operators \cite{jackiw},\cite{miransbook} can be expressed as

\begin{eqnarray}
\Gamma _{B}\left( \overline{G},\varphi _{c}\right)  &=&S\left(
\varphi
_{c}\right) -iTr\ln \overline{G}^{-1}+i\frac{1}{2}Tr\ln D^{-1}+i\frac{1}{2}%
Tr\ln \Delta ^{-1}  \nonumber \\
&&-iTr\left[ G^{-1}\left( \varphi _{c}\right) \overline{G}\right]
+\Gamma _{2}\left( \overline{G},\varphi _{c}\right) +C
\label{ef-act}
\end{eqnarray}
Here $C$ is a constant and $S\left( \varphi _{c}\right) $ is the
classical action evaluated in the scalar vev $\varphi _{c}.$
Non-bar notation indicates free propagators, as it is the case for
the gauge
\begin{equation}
D_{\mu \nu }(x-y)=\int \frac{d^{4}q}{(2\pi )^{4}}\frac{e^{iq\cdot
(x-x^{\prime })}}{q^{2}-i\epsilon }\left( g_{\mu \nu }-(1-\xi
)\frac{q_{\mu }q_{\nu }}{q^{2}-i\epsilon }\right),
\label{gaugeprop}
\end{equation}
and the scalar
\begin{equation}
\Delta (x-y)=\int \frac{d^{4}q}{(2\pi )^{4}}\frac{e^{iq\cdot
(x-x^{\prime })}}{ q^{2}+M^{2}-i\epsilon } \label{scalarprop}
\end{equation}
propagators. Here $\xi$ is the
gauge fixing parameter and $M^{2}=%
\frac{\lambda }{2}\varphi _{c}^{2}$ denotes the scalar square
mass. A dependence on full boson propagators is not included since
we do not expect the gauge field to acquire nonzero expectation
values for its composite operator. On the other hand, we are going
to explore the possibility of a non-zero vev of the scalar field,
hence, a composite-operator solution for the scalar would be a
correction of higher order that can be neglected.

 The bar on the fermion propagator $\overline{G}\left(
x,y\right) $ means that it is taken full. The full fermion
propagator in the presence of a constant magnetic field $B$ can be
written as
\cite{ackley},\cite{ferrer-incera3},\cite{ritus}-\cite{ritus-Book},

\begin{equation}
\overline{G}\left( x,y\right) =\sum\limits_{k}\int \frac{dp_{0}dp_{2}dp_{3}}{%
\left( 2\pi \right) ^{4}}E_{p}\left( x\right) \left(
\frac{1}{\gamma . \overline{p}+\Sigma (p)}\right)
\overline{E}_{p}\left( y\right). \label{fullfermionprop}
\end{equation}
with $\Sigma (p)$ being the fermion self energy,
$\overline{p}=(p_{0},0,-\sqrt{2gBk},p_{3})$, and $k$ denoting the
Landau level number. Similarly, the free fermion inverse
propagator in the presence of $B$ is given by

\begin{equation}
G^{-1}\left( x,y,\varphi _{c}\right) =\sum\limits_{k}\int \frac{%
dp_{0}dp_{2}dp_{3}}{\left( 2\pi \right) ^{4}}%
E_{p}\left( x\right) \left( \gamma .\overline{p}+\lambda
_{y}\varphi _{c}\right) \overline{E}_{p}\left( y\right).
\label{e9}
\end{equation}
Note that $\lambda _{y}\varphi _{c}$ enters as a contribution to
the fermion mass due to the shift $\varphi \rightarrow \varphi
+\varphi _{c}$ in the scalar field done in the classical action to
account for a possible non-zero scalar vev. The value of $\varphi
_{c}$ will be determined self-consistently through Eq.
(\ref{min2}).

In the above equations, Ritus' $E_{p}$ functions \cite
{ritus}-\cite{ritus-Book} were introduced. They form an
orthonormal and complete set of matrix functions and provide an
alternative method to the Schwinger's approach to problems of QFT
on electromagnetic backgrounds\footnote{For details on the use of
Ritus' method in the theory given by Eq.~(\ref{lag}) see
Refs.\cite{ferrer-incera2}-\cite{ferrer-incera1}.}. Ritus'
approach was originaly developed for spin-1/2 charged particles
\cite{ritus}-\cite{ritus-Book}, and it has been recently extended
to the spin-1 charged particle case \cite{ferrer}.

The function $\Gamma _{2}\left( \overline{G},\varphi _{c}\right) $
in (\ref {ef-act}) represents the sum of two- and higher- loop
two-particle irreducible vacuum diagrams with respect to fermion
lines. For weakly coupling theories, like the case of Lagrangian
(\ref{lag}), one can use the
Hartree-Fock approximation, which means to retain only the contributions to $%
\Gamma _{2}$ that are lowest-order in coupling constants (i.e.
two-loop graphs only), so that it becomes

\begin{eqnarray}
\Gamma _{2}\left( \overline{G},\varphi _{c}\right)
&=&\frac{e^{2}}{2}\int d^{4}xd^{4}ytr\left[\overline {G}\left(
x,y\right) \gamma ^{\mu }\overline{G}
\left( y,x\right) \gamma ^{\nu }D_{\mu \nu }(x,y)\right]  \nonumber \\
&&-\frac{e^{2}}{2}\int d^{4}xd^{4}ytr\left( \gamma ^{\mu
}\overline{G}\left( x,x\right) \right) D_{\mu \nu }(x-y)tr\left(
\gamma ^{\nu }\overline{G}
\left( y,y\right) \right)  \nonumber \\
&&+\frac{\lambda _{y}^{2}}{2}\int d^{4}xd^{4}ytr\left[
\overline{G}\left(
x,y\right) \overline{G}\left( y,x\right) \Delta (x,y)\right]  \nonumber \\
&&-\frac{\lambda _{y}^{2}}{2}\int d^{4}xd^{4}ytr\left(
\overline{G}\left( x,x\right) \right) \Delta (x-y)tr\left(
\overline{G}\left( y,y\right) \right) \label{gamma2}
\end{eqnarray}

As discussed above, the infrared dynamics ($p<<\sqrt{2eB})$ of a
system of interacting fermions in the presence of a magnetic field
is mainly governed by the contribution of the LLL
\cite{mira-gus-sho}-\cite{Gusynin}. To obtain an explicit form for
Eqs. (\ref{min1})-(\ref{min2}), we use the propagators
(\ref{gaugeprop})-(\ref{fullfermionprop}) in Eqs. (\ref{ef-act})
and  (\ref{gamma2}), and take into account that in the background
magnetic field the self-energy structure entering in the full
fermion propagator (\ref{fullfermionprop}) should be written as
\cite{ferrer-incera3}
\begin{equation}
\widetilde{\Sigma }(p)=Z_{_{\Vert }}(\overline{p})\gamma .\overline{p}%
_{_{\Vert }}+Z_{\bot }(\overline{p})\gamma .\overline{p}_{_{\bot }}+\Sigma (%
\overline{p}). \label{struct-sigma}
\end{equation}
Here we are using the notation $p_{_{\Vert }}=(p_{0},p_{3})$ and
$p_{\bot }=(p_{1},p_{2})$ for the momentum components. The wave
function renormalization coefficients $Z_{_{\Vert }}$,$Z_{\bot
}$are scalar functions of the momentum. Using this structure for
$\Sigma $ in the full fermion propagator, evaluating at the LLL
(k=0), and using the solution of the wave function
renormalization, $Z_{_{\Vert }}=0$, found in Ref.\cite
{ferrer-incera2}, we have that the gap equation (\ref{min1}) and
the scalar minimum equation (\ref{min2}) of our theory take the
form
\begin{eqnarray}
\widehat{\Sigma }(p) &=&2e^{2}\int\limits_{0}^{\infty }\frac{d^{2}\widehat{q}%
_{_{\Vert }}d^{2}\widehat{q}_{_{_{\bot }}}}{\left( 2\pi \right) ^{4}}\frac{%
\widehat{\Sigma }(\left( \widehat{q}-\widehat{p}\right) _{_{\Vert }}^{2})}{\left( \widehat{q}-%
\widehat{p}\right) _{_{\Vert }}^{2}+\widehat{\Sigma }^{2}(\left( \widehat{q}-%
\widehat{p}\right) _{_{\Vert }}^{2})}\frac{e^{-\widehat{q}_{\bot }^{2}}}{%
\widehat{q}^{2}}+\lambda _{y}^{2}\int\limits_{0}^{\infty }\frac{d^{2}%
\widehat{q}_{_{\Vert }}d^{2}\widehat{q}_{_{_{\bot }}}}{\left( 2\pi
\right) ^{4}}\frac{\widehat{\Sigma }(\left(
\widehat{q}-\widehat{p}\right) _{_{\Vert }}^{2})}{\left(
\widehat{q}-\widehat{p}\right) _{_{\Vert }}^{2}+\widehat{\Sigma
}^{2}(\left( \widehat{q}-\widehat{p}\right) _{_{\Vert
}}^{2})}\frac{e^{-\widehat{q}_{\bot
}^{2}}}{\widehat{q}^{2}+\widehat{M}^{2}}  \nonumber \\
&&-\frac{\lambda _{y}^{2}}{4\pi ^{3}}\frac{1}{{\widehat{M}^{2}}}%
\int\limits_{0}^{\infty }d^{2}\widehat{q}_{_{\Vert }}\frac{\widehat{\Sigma }(%
\widehat{q}_{_{\Vert }}^{2})}{\widehat{q}_{_{\Vert }}^{2}+\widehat{\Sigma }%
^{2}(\widehat{q}_{_{\Vert }}^{2})}+\lambda _{y}\widehat{\varphi
}_{c} \label{gapinit}
\end{eqnarray}

and

\begin{equation}
\widehat{\varphi }_{c}^{3}=\frac{3\lambda _{y}}{2\lambda \pi ^{3}}%
\int\limits_{0}^{\infty }d^{2}\widehat{q}_{_{\Vert }}\frac{\widehat{\Sigma }(%
\widehat{q}_{_{\Vert }}^{2})}{\widehat{q}_{_{\Vert }}^{2}+\widehat{\Sigma }%
^{2}(\widehat{q}_{_{\Vert }}^{2})}  \label{sca-min1}
\end{equation}
respectively. Dimensionless field-normalized quantities are
denoted by $\widehat{Q}=\frac{Q}{\sqrt{2eB}}$. Notice that if we
set $\lambda _{y}=0$ in the above equations, Eq. (\ref{gapinit})
reduces to the same gap equation found in \cite{Gusynin} for
(3+1)-dimensional QED, since, in the absence of a Yukawa term, the
theory (\ref{lag}) becomes equivalent to a QED theory on which an
extra, but disconnected, real scalar field has been added.

Changing $q_{_{\Vert }}$ to polar coordinates $\left( k,\theta
\right) $ in the above integrals and integrating in the angle, we
find
\begin{eqnarray}
\widehat{\Sigma }(p) &=&\frac{\alpha }{2\pi }\int\limits_{0}^{\infty }d%
\widehat{k}^{2}\frac{\widehat{\Sigma }(\widehat{k}^{2})}{\widehat{k}^{2}+%
\widehat{\Sigma }^{2}(\widehat{k}^{2})}{\varkappa }_{0}(\widehat{p}%
^{2},\widehat{k}^{2})+\frac{\lambda _{y}^{2}}{16\pi ^{2}}\int\limits_{0}^{%
\infty }d\widehat{k}^{2}\frac{\widehat{\Sigma }(\widehat{k}^{2})}{\widehat{k}%
^{2}+\widehat{\Sigma }^{2}(\widehat{k}^{2})}{\varkappa }_{\widehat{M}%
^{2}}(\widehat{p}^{2},\widehat{k}^{2})+  \nonumber \\
&&-\frac{\lambda _{y}^{2}}{4\pi ^{2}}\frac{1}{{\widehat{M}^{2}}}%
\int\limits_{0}^{\infty }d\widehat{k}^{2}\frac{\widehat{\Sigma }(\widehat{k}%
^{2})}{\widehat{k}^{2}+\widehat{\Sigma }^{2}(\widehat{k}^{2})}+\lambda _{y}%
\widehat{\varphi }_{c},  \label{gap}
\end{eqnarray}
\begin{equation}
\widehat{\varphi }_{c}^{3}=\frac{3\lambda _{y}}{2\lambda \pi ^{2}}%
\int\limits_{0}^{\infty }d\widehat{k}^{2}\frac{\widehat{\Sigma }(\widehat{k}^{2})%
}{\widehat{k}^{2}+\widehat{\Sigma }^{2}(\widehat{k}^{2})}
\label{sca-min}
\end{equation}
The functions $\varkappa _{t}(\widehat{p}^{2},x)$ are defined by

\begin{equation}
{\varkappa}_{t}(\widehat{p}^{2},x)=\int\limits_{0}^{\infty }dz\frac{%
e^{-z}}{\sqrt{(z+\widehat{p}^{2}+x+t)^{2}-4x\widehat{p}^{2}}}
\label{kappa-gen}
\end{equation}
To make the calculation more manageable, it is convenient to
divide the momentum integration in Eq. (\ref{gap}) in the two
regions separated by the dimensionless momentum square
$\widehat{p}^{2}$. Expanding the kernels $\varkappa
_{t}(\widehat{p}^{2},x)$ appropriately on each region, we find

\begin{eqnarray}
\widehat{\Sigma }(p) &=&\frac{\alpha }{2\pi }\left\{ \int\limits_{0}^{%
\widehat{p}^{2}}d\widehat{k}^{2}\frac{\widehat{\Sigma }(\widehat{k}^{2})}{%
\widehat{k}^{2}+\widehat{\Sigma }^{2}(\widehat{k}^{2})}\int\limits_{0}^{%
\infty }dz\frac{e^{-z}}{\widehat{p}^{2}+z}+\int\limits_{\widehat{p}%
^{2}}^{\infty }d\widehat{k}^{2}\frac{\widehat{\Sigma }(\widehat{k}^{2})}{%
\widehat{k}^{2}+\widehat{\Sigma }^{2}(\widehat{k}^{2})}\int\limits_{0}^{%
\infty }dz\frac{e^{-z}}{\widehat{k}^{2}+z}\right\}  \nonumber \\
&&+\frac{\lambda _{y}^{2}}{16\pi ^{2}}\left\{ \int\limits_{0}^{\widehat{p}%
^{2}}d\widehat{k}^{2}\frac{\widehat{\Sigma }(\widehat{k}^{2})}{\widehat{k}%
^{2}+\widehat{\Sigma }^{2}(\widehat{k}^{2})}\int\limits_{0}^{\infty }dz\frac{%
e^{-z}}{\widehat{p}^{2}+z+\widehat{M}^{2}}+\int\limits_{\widehat{p}%
^{2}}^{\infty }d\widehat{k}^{2}\frac{\widehat{\Sigma }(\widehat{k}^{2})}{%
\widehat{k}^{2}+\widehat{\Sigma }^{2}(\widehat{k}^{2})}\int\limits_{0}^{%
\infty }dz\frac{e^{-z}}{\widehat{k}^{2}+z+\widehat{M}^{2}}\right\}
\nonumber
\\
&&+\frac{2}{3}\lambda _{y}\widehat{\varphi }_{c}  \label{gap-exp}
\end{eqnarray}
Notice that we used Eq. (\ref{sca-min}) to combine the last two terms of Eq. (%
\ref{gap}) into the last term of Eq. (\ref{gap-exp}). The
analytical solutions of Eqs. (\ref{sca-min}) and Eq.
(\ref{gap-exp}) can be explored by converting first the non-linear
integral equation (\ref {gap-exp}) to a second order non-linear
differential equation. First, however, we must take into account
that the consistency of the LLL approximation requires to use a
momentum cutoff of order $\sqrt{2eB\text{ }}$in the momentum
integrations, and hence the infinity limit in all the integrals in
$\widehat{k}^{2}$ should be changed to 1.

One can easily see, by taking derivatives of Eq. (\ref{gap-exp})
with respect to $x\equiv\widehat{p}^{2}$ and combining them
conveniently, that the integral equation (\ref {gap-exp}) is
equivalent to the following second order differential equation

\begin{equation}
\widehat{\Sigma }^{\prime \prime }\left( x\right) -\frac{\overline{g}%
^{\prime \prime }\left( x\right) }{\overline{g}^{\prime }\left( x\right) }%
\widehat{\Sigma }^{\prime }\left( x\right) -\overline{g}^{\prime
}\left(
x\right) \frac{\widehat{\Sigma }\left( x\right) }{x+\widehat{\Sigma }%
^{2}\left( x\right) }=0  \label{gap-diff-eq}
\end{equation}
If we now differentiate (\ref{gap-exp}) and evaluate the result at
$x=0$, we obtain the following boundary condition

\begin{equation}
\left. \frac{\widehat{\Sigma }^{\prime }\left( x\right) }{\overline{g}%
^{\prime }\left( x\right) }\right| _{x=0}=0, \label{bound-cond}
\end{equation}
where
\begin{equation}
\overline{g}\left( x\right) =\frac{\alpha }{2\pi }g\left( x\right) +\frac{%
\lambda _{y}^{2}}{16\pi ^{2}}g\left( x+\widehat{M}^{2}\right),
\label{gbar}
\end{equation}
\begin{equation}
g(y)=\int\limits_{0}^{\infty }dz\frac{e^{-z}}{z+y}.  \label{g}
\end{equation}
Similarly, taking the derivative of (\ref{gap-exp}), multiplying
it by $\frac{\overline{g}(x)}{\overline{g}'(x)}$ and evaluating at
$x=1$, we obtain the second independent boundary condition

\begin{equation}
\left[
\widehat{\Sigma }\left( x\right) -\frac{\overline{g}\left( x\right) }{%
\overline{g}^{\prime }\left( x\right) }\widehat{\Sigma }^{\prime
}\left( x\right)\right] _{x=1}=\frac{2}{3}\lambda
_{y}\widehat{\varphi }_{c}. \label{bound-cond-x1}
\end{equation}

In doing so, we have traded a non-linear integral equation for a
non-linear boundary value problem. Finding the solutions to the
coupled set of Eqs. (\ref{sca-min}) and (\ref{gap-diff-eq}), with
boundary conditions (\ref{bound-cond}) and (\ref{bound-cond-x1})
will be the aim of the next section.

\section{Fermion and Scalar Masses in the Beyond-Constant-Mass Approximation}
\subsection{Beyond-Constant-Mass Analytical Solutions}
An analytical expression for the solution $\widehat{\Sigma }(x)$
of (\ref {gap-diff-eq})-(\ref{bound-cond}) can be found
considering a linearized version of the equations
(\ref{gap-diff-eq}) and (\ref{sca-min}), on which the fermion self
energy in the denominators is replaced by its zero momentum value
$\Sigma \left( 0\right) =m$.  The consistency of such
linearization is justified if the self energy is a rapidly
decreasing function of the momentum. We will corroborate at the
end of the derivations that follow that this is indeed the case.
Then, the gap equation (\ref{gap-diff-eq}) can be written as

\begin{equation}
\widehat{\Sigma }^{\prime \prime }\left( x\right) -\frac{\overline{g}%
^{\prime \prime }\left( x\right) }{\overline{g}^{\prime }\left( x\right) }%
\widehat{\Sigma }^{\prime }\left( x\right) -\overline{g}^{\prime
}\left( x\right) \frac{\widehat{\Sigma }\left( x\right)
}{x+\widehat{m}^{2}}=0 \label{dif-eq-2}
\end{equation}
while the equation for the scalar minimum takes the form
\begin{equation}
\widehat{\varphi }_{c}^{3}=\frac{3\lambda _{y}}{2\lambda \pi ^{2}}%
\int\limits_{0}^{1}dx\frac{\widehat{\Sigma
}(x)}{x+\widehat{m}^{2}} \label{linearized-scal-min}
\end{equation}
From a physical point of view, we expect that the masses for both
fermion and scalar fields will be much smaller than the magnetic
field that induces them through the formation of a fermion
condensate. Therefore, it is reasonable to assume that
$\widehat{m}^{2}\ll1$ and $\widehat{M}^{2}\ll1$. At the end of our
calculations we must check in the obtained results the consistency
of this assumption.

Taking into account the asymptotic behaviors of the function $\overline{g}%
\left( x\right) $ in the regions:\\

$\left( 1\right) $ \ $x<<\widehat{M}^{2}<<1$

\begin{equation}
\frac{\overline{g}^{\prime \prime }\left( x\right)
}{\overline{g}^{\prime }\left( x\right) }\simeq
-\frac{1}{x}\hspace{1in}\overline{g}^{\prime }\left( x\right)
\simeq -\frac{\alpha }{2\pi }\frac{1}{x}  \label{g-asymp-1}
\end{equation}

$\left( 2\right) $ \ $\widehat{M}^{2}<<x\leq 1$

\begin{equation}
\frac{\overline{g}^{\prime \prime }\left( x\right)
}{\overline{g}^{\prime }\left( x\right) }\simeq
-\frac{1}{x}\hspace{1in}\overline{g}^{\prime
}\left( x\right) \simeq -\left( \frac{\alpha }{2\pi }+\frac{\lambda _{y}^{2}%
}{16\pi ^{2}}\right) \frac{1}{x}  \label{g-asymp-2}
\end{equation}
one ends up with a different boundary value problem at each
region. The two boundary value problems are defined by the
following equations:\\

$\left( 1\right)\ $ For  $x<<\widehat{M}^{2}<<1$

\begin{equation}
\widehat{\Sigma }^{\prime \prime }\left( x\right) +\frac{1}{x}\widehat{%
\Sigma }^{\prime }\left( x\right) +\frac{\alpha }{2\pi }\frac{\widehat{%
\Sigma }\left( x\right) }{x\left( x+\widehat{m}^{2}\right) }=0,
\label{xlessM-problem}
\end{equation}
\begin{equation}
x\widehat{\Sigma }^{\prime }\left( x\right) _{x=0}=0  \label{bc1}
\end{equation}

$\left( 2\right)\ $ For  $\widehat{M}^{2}<<x\leq 1$
\begin{equation}
\widehat{\Sigma }^{\prime \prime }\left( x\right) +\frac{1}{x}\widehat{%
\Sigma }^{\prime }\left( x\right) +\left( \frac{\alpha }{2\pi }+\frac{%
\lambda _{y}^{2}}{16\pi ^{2}}\right) \frac{\widehat{\Sigma }\left(
x\right) }{x\left( x+\widehat{m}^{2}\right) }=0,
\label{xlargM-problem}
\end{equation}
\begin{equation}
\widehat{\Sigma }\left( 1\right) +\epsilon \widehat{\Sigma
}^{\prime }\left( 1\right) =\frac{2}{3}\lambda
_{y}\widehat{\varphi }_{c}  \label{bc2}
\end{equation}
where $\epsilon =\frac{g(1)}{\overline{g}^{\prime }(1)}=1.477$ and $\alpha =%
\frac{1}{137}$ is the fine-structure constant.

For most  physically interesting applications of the gauge Yukawa
theory, the Yukawa coupling $\lambda _{y}$ is $\leq
10^{-1}$.  For those $%
\lambda _{y}'s$, the parameters $\nu =\sqrt{\frac{\alpha }{2\pi }}$ and $%
\overline{\nu }=\sqrt{\frac{\alpha }{2\pi }+\frac{\lambda
_{y}^{2}}{16\pi ^{2}}}$ practically coincide (for $\lambda
_{y}=10^{-2}$ they already have three significant common figures).
Thus, we can take $\overline{\nu }\simeq \nu ,$ in Eq. (\ref
{xlargM-problem}), reducing the problem to a single second order
differential equation. The new problem is then defined by Eq.
(\ref {xlessM-problem}) and the two boundary conditions
(\ref{bc1}) and (\ref{bc2}). The solution to this boundary value
problem can be written as the following combination of
hypergeometric functions (for properties and formulas of the
hypergeometric functions see \cite{bateman})

\begin{equation}
\widehat{\Sigma }\left( x\right) =A_{1}F\left( i\nu ,-i\nu ;1;-\frac{x}{%
\widehat{m}^{2}}\right) +A_{2}\left(
1+\frac{x}{\widehat{m}^{2}}\right) F\left( 1-i\nu ,1+i\nu
;2;1+\frac{x}{\widehat{m}^{2}}\right)  \label{sigma}
\end{equation}
Taking into account the boundary condition (\ref{bc1}) and the
formula
\begin{equation}
\frac{dF\left( a,b;c;z\right) }{dz}=\frac{ab}{c}F\left(
a+1,b+1;c+1;z\right) \label{hyperg-deriv}
\end{equation}
we obtain $A_{2}=0.$ As $\widehat{m}=\widehat{\Sigma }\left(
0\right) ,$ it is clear that $A_{1}=\widehat{m}.$ Therefore the
self-energy solution becomes

\begin{equation}
\widehat{\Sigma }\left( x\right) =\widehat{m}F\left( i\nu ,-i\nu ;1;-\frac{x}{%
\widehat{m}^{2}}\right) \label{sigmasol}
\end{equation}
The second boundary condition (\ref{bc2}) gives rise to
\begin{equation}
\widehat{m}F\left( i\nu ,-i\nu
;1;-\frac{1}{\widehat{m}^{2}}\right)
-\epsilon \frac{\nu ^{2}}{\widehat{m}}F\left( 1+i\nu ,1-i\nu ;2;-\frac{1}{%
\widehat{m}^{2}}\right) =\frac{2}{3}\lambda _{y}\widehat{\varphi
}_{c}, \label{b2-eval}
\end{equation}
which establishes a relation between the fermion dynamical mass $\widehat{m}%
, $ and the scalar vev $\widehat{\varphi }_{c}$. This is an
implicit, quite non-trivial equation for $\widehat{m}:$
besides the dependence on $ \widehat{m}$ in the hypergeometric
functions, the scalar vev  $\widehat{\varphi }_{c}$ depends on $
\widehat{m}$  through Eq. (\ref{linearized-scal-min}).

To find the solution to the system formed by
(\ref{linearized-scal-min}) and (\ref{b2-eval}), we first note that
Eq. (\ref{xlessM-problem}) can be rewritten in the form
\begin{equation}
\frac{d}{dx}\left( x\widehat{\Sigma }^{\prime }\left( x\right)
\right) =-\nu ^{2}\frac{\widehat{\Sigma }\left( x\right)
}{x+\widehat{m}^{2}}, \label{deriv-sigm}
\end{equation}
hence
\begin{equation}
\int\limits_{0}^{1}dx\frac{\widehat{\Sigma }(x)}{x+\widehat{m}^{2}(x)}=-%
\frac{1}{\nu ^{2}}\widehat{\Sigma }^{\prime }\left( 1\right).
\label{integ-relat}
\end{equation}
Using (\ref{sigma}) and (\ref{integ-relat}) in Eq.
(\ref{linearized-scal-min}%
), and the values of $A_{1\text{ }}$and $A_{2}$ just found, we obtain
\begin{equation}
\widehat{\varphi }_{c}^{3}=\frac{3\lambda _{y}}{2\lambda \pi ^{2}}\frac{1}{%
\widehat{m}}F\left( 1+i\nu ,1-i\nu
;2;-\frac{1}{\widehat{m}^{2}}\right) \label{phi3}
\end{equation}

From the asymptotic behavior of the hypergeometric function for
large values of its argument \cite {bateman}
\begin{equation}
F\left( a,b;c;z\right) \simeq \frac{\Gamma \left( c\right) \Gamma
\left( b-a\right) }{\Gamma \left( b\right) \Gamma \left(
c-a\right) }\left(
-z\right) ^{-a}+\frac{\Gamma \left( c\right) \Gamma \left( a-b\right) }{%
\Gamma \left( a\right) \Gamma \left( c-b\right) }\left( -z\right)
^{-b} \label{largez-hyperg}
\end{equation}
we can show that
\begin{eqnarray}
F\left( i\nu ,-i\nu ;1;-\frac{1}{\widehat{m}^{2}}\right) &\simeq &\frac{%
\Gamma \left( -2i\nu \right) }{\Gamma \left( -i\nu \right) \Gamma
\left(
1-i\nu \right) }\left( \frac{1}{\widehat{m}^{2}}\right) ^{-i\nu }+\frac{%
\Gamma \left( 2i\nu \right) }{\Gamma \left( i\nu \right) \Gamma
\left( 1+i\nu \right) }\left( \frac{1}{\widehat{m}^{2}}\right)
^{i\nu }  \nonumber
\\
&\simeq &Re\left[ \frac{\Gamma \left( 1+2i\nu \right) }{\Gamma
^{2}\left( 1+i\nu \right) }\left( \frac{1}{\widehat{m}^{2}}\right)
^{i\nu }\right]
\nonumber \\
&\simeq &Re\left[ \left| \frac{\Gamma \left( 1+2i\nu \right)
}{\Gamma
^{2}\left( 1+i\nu \right) }\right| e^{i\left( \nu \ln \frac{1}{\widehat{m}%
^{2}}+\delta \left( \nu \right) \right) }\right]  \label{hyper1}
\end{eqnarray}
where
\begin{equation}
\delta \left( \nu \right) =\arg \frac{\Gamma \left( 1+2i\nu
\right) }{\Gamma ^{2}\left( 1+i\nu \right) }\approx \nu
^{3}+O\left( \nu ^{5}\right) \label{deltanu}
\end{equation}
and
\begin{equation}
\left| \frac{\Gamma \left( 1+2i\nu \right) }{\Gamma ^{2}\left(
1+i\nu \right) }\right| \approx 1,
\end{equation}
so the function can be approximated by
\begin{equation}
F\left( i\nu ,-i\nu ;1;-\frac{1}{\widehat{m}^{2}}\right) \simeq
Re\left[e^{i\nu\ln\frac{1}{\widehat{m}^{2}}}\right]=\cos \left(
\nu \ln \frac{1}{\widehat{m}^{2}}\right) \label{largez-simp1}
\end{equation}
Similarly, one can see that

\begin{equation}
F\left(1+i\nu ,1-i\nu ;2;-\frac{1}{\widehat{m}^{2}}\right) \simeq \frac{%
\widehat{m}^{2}}{\nu}\sin \left( \nu \ln
\frac{1}{\widehat{m}^{2}}\right) \label{largez-simp2}
\end{equation}
Substituting with Eqs. (\ref{largez-simp1}) and
(\ref{largez-simp2}) into (\ref{b2-eval}), (\ref{phi3}), we obtain
a much more simplified, though still transcendental, pair of
coupled equations for the fermion infrared mass $\widehat{m}$ and
the scalar vev (or equivalently, for the fermion infrared mass and
the scalar mass $\widehat{M}$),

\begin{equation}
e^{-\frac{{\verb|t|}}{\nu}}[\cos\left({\verb|t|} \right)-\nu\epsilon\sin%
\left( \verb|t|\right)]^{3}-\frac{4}{9}\frac{\lambda_{y}^{4}}{%
\pi^{2}\lambda\nu}\sin\left( \verb|t|\right)=0
\label{fermion-m-final}
\end{equation}

\begin{equation}
\widehat{M}^{2}=\frac{\lambda \widehat{\varphi
}_{c}^{2}}{2}=\frac{9\lambda }{8\lambda
_{y}^{2}}e^{-\frac{\verb|t|}{\nu }}[\cos \left( \verb|t|\right)
-\nu \epsilon \sin \left( \verb|t|\right) ]^{2},
\label{boson-m-final}
\end{equation}
where the parameter $\verb|t|=\nu \ln
(\frac{1}{\widehat{m}^{2}})$.

Eqs. (\ref{fermion-m-final})-(\ref{boson-m-final}) represents the
BCMA implicit solution  for the fermion and scalar masses
catalyzed by the magnetic field. This is as far as we can stretch
our analytical calculations for $\widehat{m}^{2}$ and
$\widehat{M}^{2}$ without introducing any additional
approximation. In the following subsections we will perform a
numerical analysis of these solutions.

\subsection{Numerical Solutions in the BCMA}

Since Eqs. (\ref{fermion-m-final})-(\ref{boson-m-final}) are
highly transcendental, to obtain the explicit dependence with the
couplings of the BCMA fermion and scalar masses, we have to resort
to numerical methods.

\begin{figure}
\epsfxsize=9cm \epsfysize=7cm\epsffile{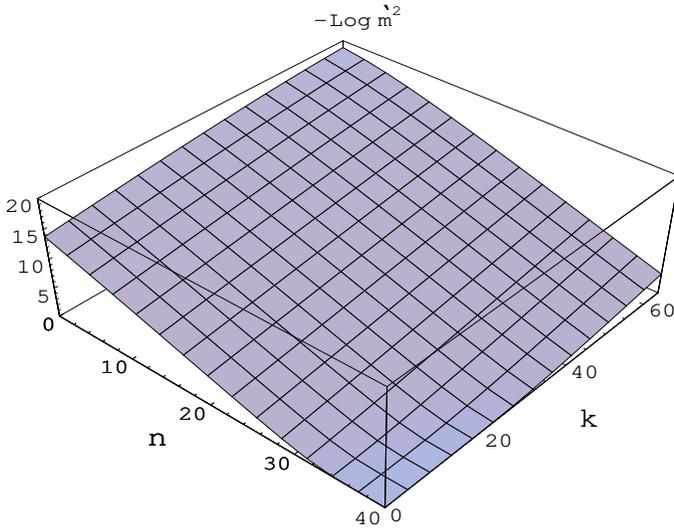}
\caption{Logarithmic plot of the BCM solution field-normalized
fermion mass square $\widehat{m}^{2}$ versus couplings
$\lambda_{y}=10^{-6+\frac{n}{8}}$ and
$\lambda=10^{-9+\frac{k}{8}}$ for $\alpha=\frac{1}{137}$. The
numbers in the x-y axes indicate the values of $n$ and $k$,
respectively. } \label{bcm-mass}
\end{figure}

\begin{figure}
\epsfxsize=10cm \epsfysize=7cm\epsffile{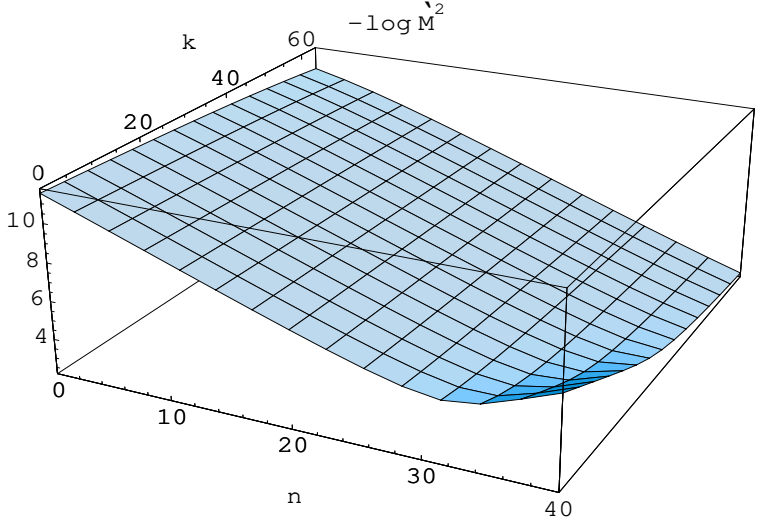}
\caption{Logarithmic plot of the dimensionless scalar square mass
$\widehat{M}^{2}$ versus Yukawa and scalar self-interaction
couplings. The intervals of the coupling constants used to
generate this graph are the same as in Fig.1.}
\label{bcm-scalmass}
\end{figure}

Figs. \ref{bcm-mass} and \ref{bcm-scalmass} display logarithmic
plots of the numerical solutions of Eqs. (\ref{fermion-m-final}),
(\ref{boson-m-final}), versus couplings $\lambda_{y}$ and
$\lambda$. From them, one can easily see that the two masses
widely agree with the initial assumptions $\widehat{m}^{2}\ll{1}$,
$\widehat{M}^{2}\ll{1}$. Only in the region of very large
$\lambda_{y}$ (very large n) and very small $\lambda$ (very small
k) the fermion mass becomes of order one, hence, to be consistent,
we should disregard the results in this corner. In any place out
of this limited section of the parameter space, the results are
reliable for both masses.

Notice that the fermion mass grows with $\lambda_{y}$ at any given
value of $\lambda$. This in turn implies an enhancement of the
fermion mass as compared to its value within QED. While in QED the
largest mass was no more than $\sim 10^{-10}\surd 2eB$
\cite{Gusynin}, here the mass surpass this value in the majority
of the parameter space in at least 5 orders of magnitude.

It is because of such a significant enhancement of the dynamically
generated mass in the presence of scalars, that the magnetic
catalysis could play an important role in realistic applications
of the HY model. The region of large $\lambda_{y}$, large
$\lambda$, where the results are quite reliable, is the most
interesting for applications to the electroweak theory, since the
values of the coupling constants in that section include the value
of the scalar self-couplings consistent with current experimental
limits for the Higgs mass, as well as the Yukawa coupling of the
top quark.

\begin{figure}
\epsfxsize=9cm \epsfysize=7cm\epsffile{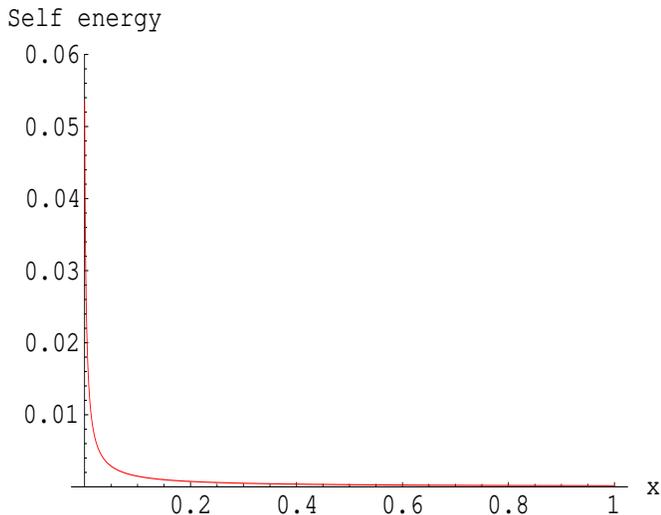}
\caption{Self energy versus momentum squared for
$\lambda_{y}=10^{-1}$ and $\lambda=10^{-2}$} \label{fig:self-en}
\end{figure}

To finish this subsection, let us consider the behavior of the
self energy with the momentum. In Fig. \ref{fig:self-en} we have
plotted the self energy solution (\ref{sigmasol}) as a function of
the momentum for fixed values of the couplings. As can be seen,
$\sum$ decreases very quickly with the momentum. This behavior is
in good agreement with the linearization used in
Eq.(\ref{dif-eq-2}). It also justifies the ultraviolet cutoff at
$\surd 2eB$ that was imposed on the integrals appearing in the gap
equation (\ref{gap-exp}) and the scalar minimum (\ref{sca-min}),
since, as seen here, the main contribution to the integrals comes
from the deep infrared region.

\subsection{Comparison Between the BCMA and the CMA Solutions}

To find the region of reliability of the CMA we have to determine
the values of the couplings for which the two approximations give
rise to the same mass solutions. With this aim we compare the BCMA
equations (\ref{fermion-m-final})-(\ref{boson-m-final}) with the
corresponding CMA equations

\begin{equation}
\widehat{m}=\frac{2}{3}\lambda _{y}\varphi _{c}+\frac{\nu
^{2}}{2}\widehat{m}\ln ^{2}\frac{1}{\widehat{m}^{2}}
\label{fermion-m-cma}
\end{equation}

\begin{equation}
\widehat{\varphi}^{3}_{c}=\frac{3\lambda_{y}}{2{\pi}^{2}\lambda}\widehat{m}
\ln \frac{1}{\widehat{m}^{2}}, \label{boson-m-cma}
\end{equation}
that were previously found\footnote{Here we have corrected some
missprints appearing in Ref. \cite{ferrer-incera1}.} in Ref.
\cite{ferrer-incera1}.

Eqs. (\ref{fermion-m-cma})-(\ref{boson-m-cma}) look very different
from their BCMA counterpart: Eqs.
(\ref{fermion-m-final})-(\ref{boson-m-final}). There is no reason
to anticipate that the solutions of both sets of equations will
coincide in all the parameter space. Combining Eqs.
(\ref{fermion-m-cma})-(\ref{boson-m-cma}) we obtain

\begin{equation}
\frac{1}{\widehat{m}^{2}}\ln\frac{1}{\widehat{m}^{2}}=
\frac{9\lambda\pi^{2}}{4\lambda^{4}_{y}}
[1-\frac{1}{2}\overline{\nu}^{2}ln^{2}\frac{1}{\widehat{m}^{2}}]^{3},
\label{fermion-m-cma2}
\end{equation}

From (\ref{fermion-m-cma2}) we see that since $\widehat{m}^{2}$
has to be positive, the consistency of the CMA solution requires
$\frac{1}{2}\overline{\nu}^{2}ln^{2}\frac{1}{\widehat{m}^{2}}<1$,
which is equivalent to have  $t<1.4$. Below, we will numerically
check that this condition is indeed always satisfied.

To compare the BCMA and CMA solutions we will explore whether
there is a condition under which the CMA and BCMA equations reduce
to an identical set. To this end, let us assume that
$\overline{\nu}ln\frac{1}{\widehat{m}^{2}} \simeq\nu
ln\frac{1}{\widehat{m}^{2}}\ll 1$. This restriction allows us to
write Eqs. (\ref{fermion-m-final})-(\ref{boson-m-final}) as
\begin{equation}
\frac{1}{\widehat{m}^{2}}\ln(\frac{1}{\widehat{m}^{2}})=\frac{9\pi^{2}\lambda%
}{4\lambda_{y}^{4}},  \label{rough-m}
\end{equation}
\begin{equation}
\widehat{M}^{2}=\frac{9\lambda}{8\lambda_{y}^{2}}\widehat{m}^{2}
\label{rough-M}
\end{equation}
respectively. They are exactly the same equations found from
(\ref{boson-m-cma}) and (\ref{fermion-m-cma2}), after using
$t\ll1$. Thus, in this limiting case, the BCMA reduces to the CMA,
thereby $t\ll1$ defines a condition of reliability of the CMA.

The explicit region of parameter space where the CMA is reliable
can be determined from a numerical plot of the ratio between the
CMA and BCMA mass square solutions. To be sure that we are working
with consistent masses, we will restrict the couplings to a strip
in the ($\lambda_{y}$,$\lambda$) plane, leaving out the corner of
Fig. \ref{bcm-mass}, where, as discussed above, the consistency of
the approximation breaks down.

\begin{figure}
\epsfxsize=9cm\epsfysize=7cm\epsffile{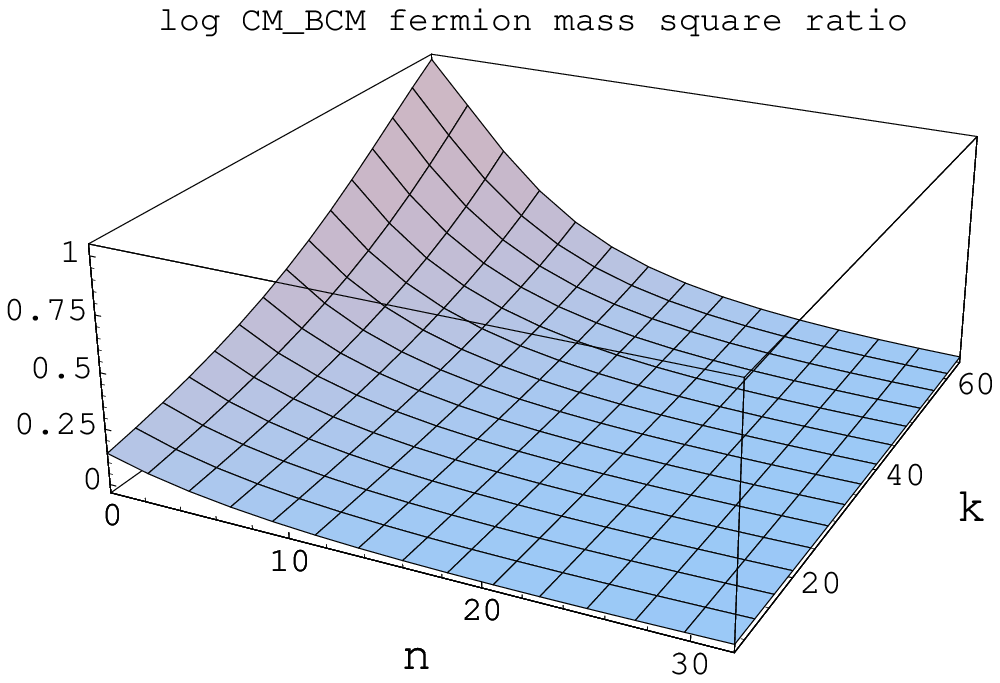}
\caption{ $Log
(\frac{\widehat{m}_{CM}^{2}}{\widehat{m}_{BCM}^{2}})$ in the
region of couplings $10^{-8}<\lambda<10^{-1}$,
$10^{-6}<\lambda_{y}<10^{-2}$.} \label{bcm-cm-fermass}
\end{figure}

\begin{figure}
\epsfxsize=9cm\epsfysize=7cm\epsffile{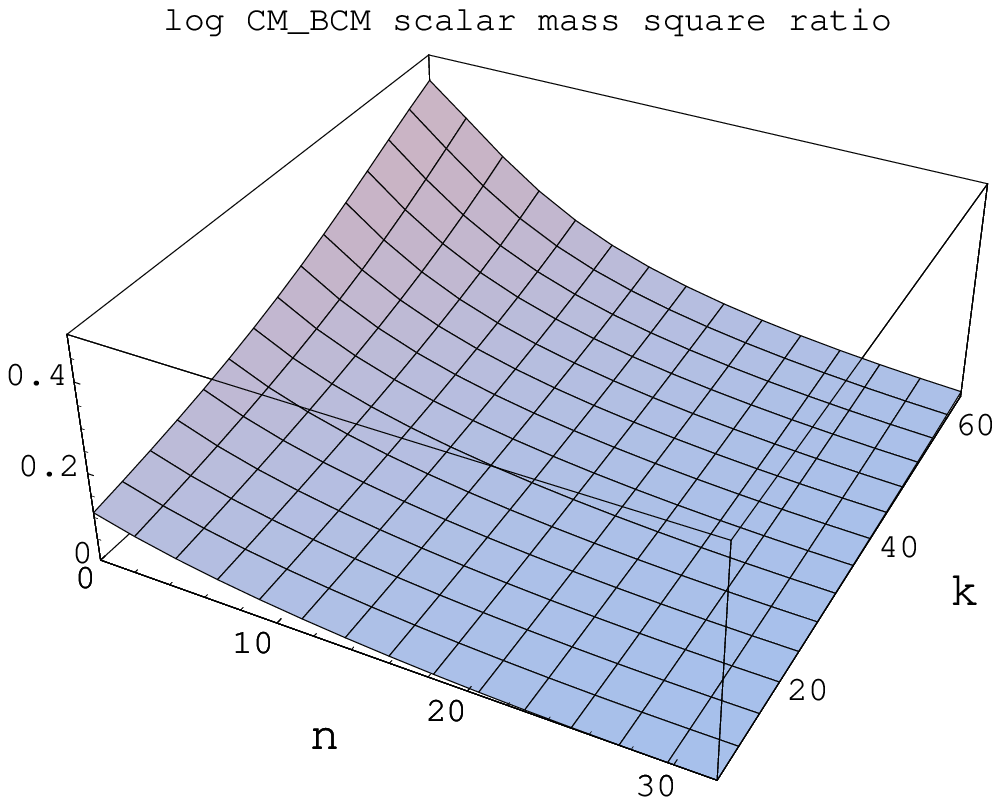}
\caption{ $Log
(\frac{\widehat{M}_{CM}^{2}}{\widehat{M}_{BCM}^{2}})$ in the
region of couplings $10^{-8}<\lambda<10^{-1}$,
$10^{-6}<\lambda_{y}<10^{-2}$.} \label{bcm-cm-scalmass}
\end{figure}

Figs. \ref{bcm-cm-fermass} and \ref{bcm-cm-scalmass} show
logarithmic plots of the ratio of CMA over BCMA mass square
results for fermion and scalar masses respectively, taken in the
region of couplings $10^{-8}<\lambda<10^{-1}$,
$10^{-6}<\lambda_{y}<10^{-2}$. Both figures display similar
behavior of the ratios, characterized by a discernible region of
the parameter space, approximately given by
$10^{-4}<\lambda<10^{-1}$ and $10^{-6}<\lambda_{y}<10^{-5}$, where
a disagreement between BCMA and CMA results is apparent. However,
even in this segment, the BCMA and CMA  mass squares differ at
most in one order of magnitude. Out of this limited region we find
very good agreement between BCMA and CMA results, particularly at
large $\lambda_{y}$, indicating that this is the most reliable
region of the CMA solution within this model.

\begin{figure}
\epsfxsize=9cm\epsfysize=7cm\epsffile{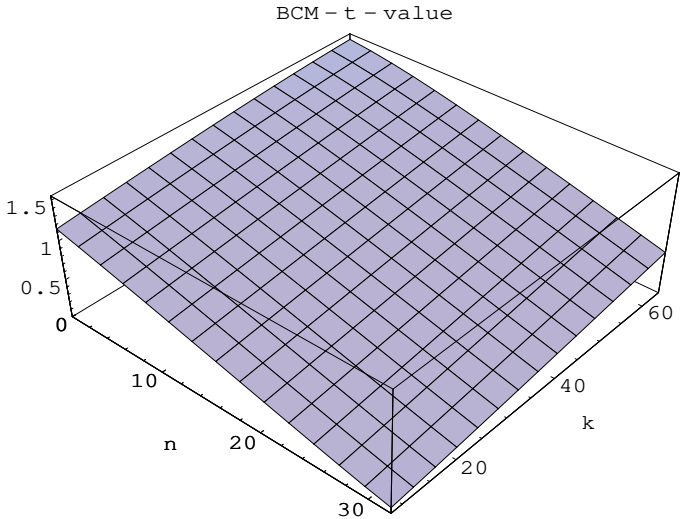}
\caption{The BCM-t-value as a function of
$\lambda_{y}=10^{-6+\frac{n}{8}}$ and
$\lambda=10^{-9+\frac{k}{8}}$.} \label{bcm-t-value}
\end{figure}

\begin{figure}
\epsfxsize=9cm\epsfysize=7cm\epsffile{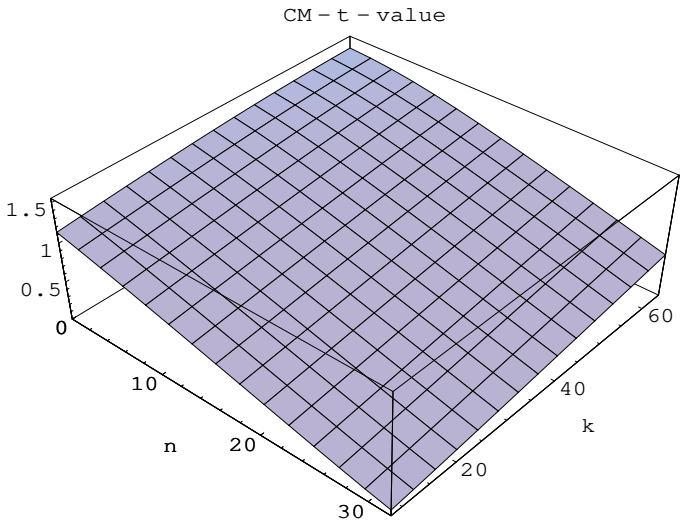}
\caption{The CM-t-value as a function of
$\lambda_{y}=10^{-6+\frac{n}{8}}$ and
$\lambda=10^{-9+\frac{k}{8}}$.} \label{cm-t-value}
\end{figure}

The above observations are corroborated by the plots of the BCM-
and CM- $t$'s, as shown in Figs. \ref{bcm-t-value} and
\ref{cm-t-value} respectively. Both surfaces have similar t-values
at equal set of couplings, even when $t\ll 1$ is not satisfied,
indicating that, after all, and as already seen in Figs.
\ref{bcm-cm-fermass} and \ref{bcm-cm-scalmass}, the two
approximations give rise to very near mass values. Notice that the
larger the $\lambda_{y}$'s, the smaller the $t$'s in both
approximation, leading to a better agreement between BCMA and CMA
results, as expected from our previous analytical considerations.
Therefore, although the numerical calculations show that the CMA
results are widely reliable, it is in this extreme section of the
parameter space where the two approximations totally coincide.
From Fig. \ref{cm-t-value} it is evident that the CM-$t$ never
goes over the limiting value of 1.4, so, even in the region of
larger discrepancy between the CM and BCM results (large
$\lambda$, relatively low $\lambda_{y}$), the CM mass solution
remains real, as it should. The curves reflect the fact that the
CM-approximation tends to overestimate the mass, because it
substitutes in the integrals the self-energy function, which
rapidly decreases with momentum, by a constant.

\subsection{BCMA in the $\lambda_{y}=0$ Limit (QED case)}

We shall discuss now the limiting case $\lambda_{y}=0$ which
reduces to (3+1)-dimensional QED with a decoupled self-interacting
scalar field. Let us find the solutions for the masses in this
case. It is clear from Eq. (\ref{sca-min}) that no scalar vev, and
hence no scalar mass, is generated in this case. The fermion
dynamical mass solution can be found from (\ref{fermion-m-final})
evaluated at $\lambda_{y}=0$. It leads to

\begin{equation}
\tan\left({\verb|t|} \right)=\frac{1}{\nu\epsilon}. \label{qed1}
\end{equation}
In terms of $\widehat{m}^{2}$,  it can be rewritten as follows

\begin{equation}
\widehat{m}^{2}=e^{-\frac{1}{\nu}\arctan(\frac{1}{\nu\epsilon})}
\label{qed2}
\end{equation}
Taking into account that $\nu\epsilon\ll1$ and using the
asymptotic behavior $\arctan(x)\simeq\frac{\pi}{2} -\frac{1}{x}$,
we obtain

\begin{equation}
\widehat{m}^{2}=e^\epsilon e^{-\pi\surd\frac{\pi}{2\alpha}.}
\label{qed3}
\end{equation}
This result coincides with the BCMA results found for QED within
the ladder approximation (see Ref. \cite{Gusynin} for details). As
known, it is qualitatively very close to its CMA counterpart
$\widehat{m}^{2}\simeq e^{-\pi\surd\frac{\pi}{\alpha}}$
\cite{Gusynin}. Thus, we are corroborating here the conclusion of
the authors of Ref. \cite{Gusynin}, namely, the reliability of the
CMA approach in (3+1)-QED\footnote{The agreement between CMA and
BCMA results in QED have been also proved using an improved ladder
approximation of the gap equation \cite{gms2}, on which the
one-loop photon propagator is used in the gap equation, instead of
the bare photon propagator.}.

It is worth to notice that the dynamical mass behavior is
basically affected by the infrared conditions of the self energy,
but it is practically indifferent to the ultraviolet boundary
condition used in Ref.\cite{Gusynin}. This explains why, despite
using a momentum cutoff at $\surd 2eB$ and imposing the second
boundary condition at $x=1$, we still get in the $\lambda_{y}=0$
case the same result as in \cite{Gusynin}, where the momentum was
allowed to run up to infinity.

\section{Concluding Remarks}

In this paper we have performed a BCMA study of the magnetically
catalyzed fermion and scalar masses in a (3+1)-dimensional Abelian
Higgs-Yukawa theory in the presence of a constant magnetic field.
Our results show that even in this multiple-coupling theory, the
discrepancy between the masses obtained within the CMA and within
the more accurate BCMA is not very significant, being the
difference in the mass square of at most one order of magnitude.
We find that the region where CMA and BCMA results exactly
coincide is defined by the condition $\verb|t|=\nu \ln
(\frac{1}{\widehat{m}^{2}})\ll 1$.

The BCMA calculations led to fermion masses many orders of
magnitude larger than those obtained in the QED case, thereby
confirming, within a more accurate approximation, that the Yukawa
interactions strengthen the generation of the dynamical fermion
mass by several orders of magnitude, a claim done in previous
papers  \cite{ferrer-incera2,ferrer-incera1} based only on CMA
results.

As mentioned in the Introduction, a motivation for the inclusion
of fermions-scalar interactions in the study of the magnetic
catalysis was to find out if this phenomenon could influence the
phenomenology of the early universe. A fundamental question here
to understand is whether the strengthening of the mass by the
fermion-scalar interactions may have any impact in the electroweak
phase transition. For this effect to be of any significance for
the electroweak physics, a condition has to be met: during the
electroweak transition the universe has to be permeated by a
primordial magnetic field strong enough as to induce, even at
temperatures comparable to the electroweak critical temperature, a
modification in the value of the fermion mass.

We should keep in mind that at temperatures below, but close
enough, the critical temperature for the electroweak spontaneous
symmetry breaking, the fermion masses generated through the Higgs
mechanism are very small, since the transition is expected to be
either second order or weakly first order. Then, if the magnetic
field is much larger than these tiny masses, the fermions will be
mainly constrained to their LLL and the MC can be fully operative.
However, this is only true if the thermal fluctuations are not as
important as to take the fermions out of the LLL. Another way to
put this is to say that the critical temperature at which the
magnetically induced fermion mass evaporates has to be larger than
the electroweak critical temperature.

Magnetic fields may have well been present at the early universe.
In fact, there are very plausible arguments favoring the existence
of primordial magnetic fields that can serve as the source of the
seed fields required to explain the observed magnetic fields in
galaxies and clusters of galaxies \cite{grasso}. The literature on
this topic is rich in possible primordial fields generating
mechanisms, and many of them can produce very strong fields at and
before the electroweak transition \cite{vachaspati,ambjorn}.

Although the model used in our calculations lacks the complexity
of the electroweak theory, it shares some common features with the
electromagnetic sector of the electroweak model, and as so we
expect that any conclusion drawn within our model can be seen as
an indication (even if qualitative) of what the relevance of the
effect would be in the electroweak context.

Taking into account that the critical temperature for the
vanishing of the magnetically catalyzed fermion mass is typically
of the order of the value of the dynamical mass at zero
temperature \cite{10,ferrer-incera2}, that is $T\sim
{m}_{d}(T=0)$, and that a reasonable estimate
\cite{vachaspati,ambjorn} for the primordial magnetic field at the
electroweak scale is $\sim10^{24}G$, one obtains, for the values
of $\lambda_{y}$ and $\lambda$ that gives rise to the largest
zero-temperature dynamical mass, that $T_{c}\sim {1
GeV}\ll{T}_{ew}\simeq100 GeV$. Hence, no magnetically induced mass
would be present at the electroweak temperature because
temperature effects override field effects at this scale. Unless
new sources of extremely large $B>>T^{2}$ primordial magnetic
fields can be identified in the future, these results indicate
that the MC has no relevance during the electroweak transition.

Nevertheless, the outcomes of this work may be important for
applications of the HY model in situations where magnetic field
effects are present at sufficiently low temperatures. We expect
that they will be particularly relevant in condensed matter
applications. As mentioned in the Introduction, a HY theory has
been proposed \cite{vojta,kvesh} to describe the observed
emergence of a secondary quasiparticle gap in high-$T_{c}$
superconductors at certain doping levels. According to recent
experiments \cite{dagan}, the secondary gap can be also triggered
by an applied magnetic field. The resemblance of this behavior
with the MC is intriguing and deserve a thorough investigation.
Such an study, in turn, will require the extension of the results
of the present paper to the two-dimensional case in order to make
quantitative predictions that can be compared with the experiment.

\medskip

\textbf{Acknowledgments}

The authors are grateful to V. P. Gusynin for useful discussions.
EE is indebted with the Department of Mathematics and Center for
Theoretical Physics, MIT, in special with Dan Freedman and Bob
Jaffe, for warm hospitality. EJF and VI would like to thank the
Institute for Space Studies of Catalonia and the University of
Barcelona for their warm hospitality. The work of EE was supported
in part by DGI/SGPI (Spain), project BFM2000-0810, and by CIRIT
(Catalonia), contract 1999SGR-00257. The work of EJF and VI was
supported in part by NSF grant PHY-0070986.


\begin{references}

\bibitem{mira-gus-sho}  V. P. Gusynin, V. A. Miransky, and I. A. Shovkovy,
Phys. Rev. Lett. 73 (1994) 3499; Phys. Lett. B 349 (1995) 477.

\bibitem{Gusynin}V. P. Gusynin, V. A. Miransky, and I. A. Shovkovy,
Phys. Rev. D 52 (1995) 4747; Nucl. Phys. B 462 (1996) 249.

\bibitem{Klimenko}  K.~G.~Klimenko, Z. Phys. C 54 (1992) 323.

\bibitem{hong} D. K. Hong, Phys. Rev. D 54 (1996) 7879.

\bibitem{ackley}  C. N. Leung, Y. J. Ng, and A. W. Ackley, Phys. Rev D 54
(1996) 4181; D.-S Lee, C. N. Leung, and Y. J. Ng, Phys. Rev D 55
(1997) 6504.

\bibitem{9}  I.A. Shushpanov and A. V. Smilga, Phys. Lett. B 402 (1997)
351.

\bibitem{10}  V. P. Gusynin and I. A. Shovkovy, Phys. Rev D 56 (1997) 5251.

\bibitem{hong2}D. K. Hong, Phys. Rev. D 57 (1998) 3759.

\bibitem{gms2}V. P. Gusynin, V. A. Miransky, and I. A. Shovkovy,
Nucl. Phys. B 563 (1999) 361.

\bibitem{ferrer-incera4}  V. P. Gusynin, E. J. Ferrer and V. de la Incera,
Phys. Lett. B 455 (1999) 217.

\bibitem{ferrer-incera3}  E. J. Ferrer and V. de la Incera, Phys. Rev. D 58
(1998) 065008.

\bibitem{ferrer-incera2}  E. J. Ferrer and V. de la Incera,
Int. J. Mod. Phys. 14 (1999) 3963. 

\bibitem{ferrer-incera1}  E. J. Ferrer and V. de la Incera, Phys. Lett. B
481 (2000) 287.

\bibitem{incera} V. de la Incera, Proceedings of the International Conference On
Quantization, Gauge Theory, And Strings: Conference Dedicated To
The Memory Of Professor E.S. Fradkin, pag. 316, edited by A.
Semikhatov, M. Vasiliev and V. Zaikin, Scientific World Publ.,
5-10 Jun 2000, Moscow, Russia.

\bibitem{klim}D. Ebert, V. V. Khudyakov, K. G. Klimenko, H. Toki, and V. Ch.
Zhukovsky, hep-ph/0108185;
V.C. Zhukovsky, V.V. Khudyakov, K.G. Klimenko and D. Ebert, JETP
Lett. 74 (2001) 523;
D. Ebert, V.V. Khudyakov, K.G. Klimenko and V.Ch. Zhukovsky,
Phys.Rev.D 65 (2002) 054024.

\bibitem{mira-sho} V.A. Miransky and I.A. Shovkovy, Phys.Rev.D66
(2002) 045006.

\bibitem{mira-sho-gus-2} V. P. Gusynin, V. A. Miransky, and I. A. Shovkovy,
Phys.Rev.D 67 (2003) 107703.

\bibitem{cond-mat1} G.~W.~Semenoff, I.~ A.~Shovkovy, and
L.~C.~R.~Wijewardhana, Mod. Phys. Lett. A 13 (1998) 1143; W.~V.
Liu, cond-mat/9808134; K. Farakos and N.~E.~Mavromatos, Phys. Rev.
B 57 (1998) 3017; Int. J. Mod. Phys. B 12 (1998) 2475;

\bibitem{cond-mat2} V. Ch. Zhukovsky, K. G. Klimenko, and V. V. Khudyakov, Theor.
Math. Phys. 124 (2000) 1132;
JETP Lett. 73 (2001) 121.

\bibitem{cond-mat3} E. V. Gorbar, V. P. Gusynin, V. A. Miransky and I. A.
Shovkovy, Phys.Rev. B66 (2002) 045108.

\bibitem{ferrer-incera5} V. P. Gusynin, E. J. Ferrer and V. de la Incera,
Mod.Phys.Lett.B 16 (2002) 107;
Eur. Phys. Jour. B33 (2003) 397.

\bibitem{vojta} M. Vojta, Y. Zhang, and S. Sachdev,Phys. Rev.
Lett. 85 (2000) 4940.

\bibitem{grasso} D. Grasso and H. R. Rubinstein,Phys. Rep. 348
(2001) 163.

\bibitem{durst-lee}A. C. Durst and P. A. Lee, Phys. Rev.B 62 (2000) 1270.

\bibitem{exp-1}T. Valla et al., Science 285 (1999) 2110; Phys. Rev.
Lett. 85 (2000) 828.

\bibitem{kvesh} D. V. Khveshchenko, and J. Paaske, Phys. Rev.
Lett. 86 (2001) 4672.

\bibitem{dagan} Y. Dagan and G. Deutscher,Phys. Rev.
Lett. 87 (2001) 177004.

\bibitem{alexandre}J. Alexandre, K. Farakos and G. Koutsoumbas, Phys. Rev. D
62 (2000) 105017;
Phys. Rev. D 63 (2001) 065015.

\bibitem{alexandre2}J. Alexandre, K. Farakos and G. Koutsoumbas, Phys. Rev. D
64 (2001) 067702.

\bibitem{jackiw}  J. M. Cornwall, R. Jackiw and E. Tomboulis, Phys. Rev. D
10 (1974) 2428.

\bibitem{miransbook}  V.A. Miransky, Dynamical Symmetry Breaking in QFT,
(World Scientific, Singapore, 1993).

\bibitem{ritus}  V. I. Ritus, Ann. of Phys. (NY) \textbf{69} (1972) 555;
ZhETF \textbf{75} (1978) 1560 (Sov. Phys. JETP \textbf{48} (1978)
788); ZhETF \textbf{76} (1979) 383.

\bibitem{ritus-Book}  V. I. Ritus in Issues in Intense-Field Quantum
Electrodynamics, ed. V. L. Ginzburg. Proceedings of\textit{\ }P.
N. Lebedev Physical Institute, Vol. \textbf{111}, 5 (1979, Moscow
Nauka) (Engl. Transl., V.168, Nova Science, Commack, 1987).

\bibitem{ferrer} E. Elizalde, E. J. Ferrer, and V. de la Incera,
Annals of Phys. (NY) 295 (2002) 33.

\bibitem{bateman}  A. Erderlyi, Higher Transcendental Functions, Vol. 1
(Bateman Manuscript Project, CalTec, McGraw-Hill Book Co.).

\bibitem{vachaspati} T. Vachaspati, Phys. Lett. B 265 (1991) 258.

\bibitem{ambjorn}J. Ambjorn and P. Olesen, hep-ph/9304220.

\end{references}
\end{document}